# Infrared-active excitations related to Ho$^{3+}$ ligand-field splitting at the commensurate-incommensurate magnetic phase transition in HoMn$_2$O$_5$

A. A. Sirenko,[a] S. M. O'Malley, and K. H. Ahn

Department of Physics, New Jersey Institute of Technology, Newark, New Jersey 07102

S. Park

Rutgers Center for Emergent Materials and Department of Physics and Astronomy, Rutgers University, Piscataway, New Jersey 08854

G. L. Carr

National Synchrotron Light Source, Brookhaven National Laboratory, Upton, New York 11973

S-W. Cheong

Rutgers Center for Emergent Materials and Department of Physics and Astronomy, Rutgers University, Piscataway, New Jersey 08854

---

[a] sirenko@njit.edu

## ABSTRACT

Linearly polarized spectra of far-infrared (IR) transmission in $HoMn_2O_5$ multiferroic single crystals have been studied in the frequency range between 8.5 and 105 cm$^{-1}$ and for temperatures between 5 K and 300 K. Polarization of IR-active excitations depends on the crystallographic directions in $HoMn_2O_5$ and is sensitive to the magnetic phase transitions. We attribute some of the infrared-active excitations to electric-dipole transitions between ligand-field split states of $Ho^{3+}$ ions. For light polarization along crystalline *b*-axis, the oscillator strength of electric dipoles at low frequencies (10.5, 13, and 18 cm$^{-1}$) changes significantly at the commensurate-incommensurate antiferromagnetic phase transition at $T_3 = 19$ K. This effect shows a strong correlation with the pronounced steps of the *b*-directional static dielectric function. We propose that the ligand field (LF) on $Ho^{3+}$ connects the magnetism and dielectric properties of this compound through coupling with the Mn spin structure. We comment on the possibility for composite excitations of magnons and excited LF states.

## I. INTRODUCTION

Interplay between the long-range magnetic and ferroelectric ordering has recently motivated extensive studies of rare-earth multiferroic manganites $R$Mn$_2$O$_5$ ($R$ = Tb, Dy, and Ho).[1][2][3] These materials are attractive for both fundamental studies and for possible device applications due to the intriguing phase diagram and magnetic field induced spontaneous electric polarization. A number of phase transitions occur at low temperatures. Upon cooling, Mn spins order in the *a-b* plane at $T_1 \approx 40 - 43$ K and incommensurate order of Mn spins switches to commensurate one at $T_2 \approx 37 - 39$ K and back to incommensurate order again at $T_3 \approx 15 - 25$ K. Finally, rare-earth spins order at lower temperatures below $T_4 \leq 10$ K. Anomalies in the dielectric constant and thermal expansion at $T_1$, $T_2$, $T_3$ and $T_4$ as well as appearance of spontaneous polarization along *b* axis at $T_C$, which coincides in HoMn$_2$O$_5$ with $T_2$, reflect changes of ferroelectric order and provide evidence for strong spin-lattice coupling in $R$Mn$_2$O$_5$ multiferroics.[4]

Recent optical studies of spin-lattice dynamics in multiferroics progressed in several directions: conventional Raman scattering spectroscopy of optical phonons,[5][6] infrared (IR) spectroscopy of the phonons,[7] as well as the low-frequency excitations called "electromagnons".[8][9][10] The report by Mihailova *et al.*[5] on the absence of optical phonon anomalies at the phase transitions in $R$Mn$_2$O$_5$ (Ho, Tb) was followed by the observation of a very small (< 1 cm$^{-1}$) anomalous shifts of the high frequency optical phonons around $T_1$ in $R$ = Bi, Eu, and Dy compounds.[6] While spectra of the Raman-active optical phonons demonstrate only minute variations, IR-active excitations in $R$MnO$_3$ ($R$ = Gd and Tb)[7] and TbMn$_2$O$_5$ [8,10] show intriguingly strong changes at the phase transition temperatures.

In this paper we present a systematic study of the far-IR transmission spectra in HoMn$_2$O$_5$ at low temperatures in the vicinity of the phase transitions using synchrotron radiation-based Fourier transform infrared (FT-IR) spectroscopy. We report strong changes in the oscillator strength for low-frequency IR-excitations measured for light polarization along different crystalline axes at the magnetic commensurate-incommensurate transition $T_3$ = 19 K. Weaker changes in the IR excitation spectra at around $T_1$ = 44 K were followed by a slow decrease of absorption in the temperature range much higher than $T_1$ up to about 200 K. We explain the observed IR-excitations at 10.5, 13, and 18 cm$^{-1}$ as electric-dipole transitions between ligand-field (LF) states of Ho$^{3+}$ ions. The change in the oscillator strength for low-energy IR excitations at $T_3$ is associated with modification of the wavefunction symmetry for the Ho$^{3+}$ LF states, which is influenced by magnetic ordering of the Mn spin system.

## II. EXPERIMENT

The high-temperature flux growth technique was utilized to produce bulk crystals of HoMn$_2$O$_5$ (see Ref. [1] for details). Samples with the thickness of about 0.5 mm were oriented using x-ray diffraction. The opposite sides were polished and wedged with a ~7° offset in order to suppress interference fringes. Transmission intensity was measured at the U12IR beamline at the National Synchrotron Light Source at Brookhaven National Laboratory equipped with an Oxford optical cryostat, Bruker 125HR spectrometer, LHe-pumped (~1.4 K) bolometer, and two grid polarizers for incident and transmitted radiation, respectively. The frequency resolution of 0.6 cm$^{-1}$ was chosen to be about three times smaller than the typical width of the absorption lines. The transmission intensity was measured with the incident beam perpendicular to the sample plane in various polarization configurations *z(x,y)*, where *z* stays for the direction of the light propagation, *x* and *y* denote electric polarization of incident and transmitted light, respectively. For each sample the raw data of transmitted intensity were normalized to transmission through an empty aperture with a size close to that of the sample.

Figure 1 shows transmission spectra of HoMn$_2$O$_5$ measured in *a(b,b)*, *b(a,a)*, and *a(c,c)* configurations at $T = 5$ K, which are dominated by absorption lines at 10.5, 13, 18, 37, 42, 52, and 97 cm$^{-1}$. Fast fringes between 60 and 90 cm$^{-1}$ are due to the Fabry-Perot interference in the sample. A couple of additional higher-frequency absorption lines at 31 and 85 cm$^{-1}$ appear in the spectra at higher temperatures and their oscillator strengths increases quasi linearly with temperature up to about $T = 60$ K. Transmission intensity maps (Fig. 2) were measured in *a(b,b)*, *b(a,a)*, and *a(c,c)* configurations with the temperature increments of 0.5 K in the range of $T < 50$ K and about 5 K at $T > 50$ K. Drastic increase of absorption at 10.5, 13, 18 cm$^{-1}$ and 42 cm$^{-1}$ occurs upon cooling at $T_3 = 19$ K for the *a(b,b)* configuration, where electric field $\vec{e}$ of both incoming and transmitted light is parallel to the direction of spontaneous electric polarization along *b*-axis.

Comparison of the temperature-frequency maps measured for several different $z(x,y)$ configurations demonstrated a relatively weak dependence on the direction of the light propagation direction $z$ as it is expected for oscillators with the $\vec{k}$-vector $\approx 0$. At the same time, absorption spectra of anisotropic $HoMn_2O_5$ crystals depend strongly on the light *polarization* with respect to the crystallographic directions, in particular, whether we have the electric field of light perpendicular or parallel to the Mn spin ordering plane: $\vec{e} \parallel c$ and $\vec{e} \perp c$, respectively. For example, in the *a(c,c)* transmission configuration the doublet at 10.5 and 13 cm$^{-1}$ dominates the low-frequency part of the spectrum [see Fig. 2(b)]. Close to the temperature interval between $T_2 = 39$ K and $T_1 = 43$ K, this doublet merges to a single line with a frequency of about 12 cm$^{-1}$, which maintains a significant oscillator strength further above $T_1 = 43$ K, and it can be seen even at $T = 150$ K. The 18 cm$^{-1}$ oscillator is stronger in *a(b,b)*, is weaker in *b(a,a)* configurations, and is not optically active for *a(c,c)* transmission in the entire temperature range. Higher frequency IR absorption lines do not show significant changes at the phase transition temperatures. The broad feature at 97 cm$^{-1}$ is polarized in the *a-b* plane and is not active for the light polarization along *c* axis in the broad temperature range. Weaker high-frequency absorption peaks at 80 and 85 cm$^{-1}$ develop at the temperatures lower than the commensurate-incommensurate transition $T_3$ with their oscillator strength increasing quasi-linearly with temperature up to about 60 K. Although their strength is small compared to that of other oscillators, these particular high-frequency lines will be important for our interpretation of the observed IR transitions as due to the LF states of $Ho^{3+}$.

Assignment of the observed IR excitations to electric-dipole or magnetic-dipole transitions is not straightforward in transmission experiments. Additional complication arises in multiferroic crystals due to a possible mixture of magnetic- and electric activity for some excitations. In theory, such excitations can contribute to both, electric $\chi_e(\omega)$ and magnetic $\chi_m(\omega)$ susceptibilities, both of which are the complex functions of frequency. Thus, a single response function of the sample (*e. g.*, transmission) cannot be used to distinguish between the magnetic or electric dipole activity. Among more reliable experimental techniques are variable-incidence-angle reflectivity and, of course, full-Muller matrix ellipsometry.[11] Note, however, that these optical techniques are not routinely available due to several factors, such as the lack of broad-band quarter-wave plates for the far-IR spectral range and the small size of multiferroic single crystals. Here we will make an attempt to determine the dominating type of the dipole activity for observed IR excitations based on analysis of linear polarization of the transmitted intensity. The main indications of the electric-dipole activity is domination in only one polarization, like that for the doublet at 10.5 and 13 cm$^{-1}$ in both *a(c,c)* and *b(c,c)* configurations. In contrast, the peak at 97 cm$^{-1}$, which is strong in both *c(a,a) and c(b,b)* and is practically absent in *a(c,c)* configuration, corresponds to a magnetic-dipole polarized along *c* axis. Table I presents the summary for the frequency, polarization, and dipole activity for the observed IR excitations.

The low-frequency parts of the transmission spectra, which is dominated by the electric-dipole excitations, were fitted using *SCOUT-2* software based on the Lorentz model for the parametric description of the dielectric function

$$\varepsilon(\omega) = \varepsilon_\infty + \sum_j^N \frac{S_j \Omega_j^2}{\Omega_j^2 - \omega^2 - i\gamma_j \omega}, \qquad (1)$$

where $\Omega_j$, $S_j$, and $\gamma_j$ are the oscillator's frequency, strength, and damping, respectively. Oscillator parameters for the primary absorption lines are summarized in Table 1. Since the oscillator frequencies $\Omega_j$ are practically constant for all excitations in the temperature range $T \leq 70$ K, it is convenient to present the oscillator strength $S_j$ in the same units as that for the static dielectric function by adding $\Omega^2_j$ in the numerator of Eq. (1). The primary changes in the IR spectra at the magnetic phase transitions revealed themselves as an increase in the oscillator strengths but not as the frequency softening as it is usually observed in ferroelectrics due to the Lyddane-Sachs-Teller relation.[12] Figure 3 shows a comparison between the static dielectric function $\varepsilon(T)$ and oscillator strengths $S_j(T)$ for electric-dipole active excitations. We can quantitatively relate the step-like behavior of the *b*-directional static dielectric function $\varepsilon_b(T)$ at $T_3$ with the contribution of these particular excitations (see Fig. 3(a) and Refs. [2,4] for details of dielectric measurements). In *a(b,b)* polarization, the low-frequency doublet at 10.5 and 13 cm$^{-1}$ provides the strongest contribution of about 2.2 to $\varepsilon_b(T)$ [Fig. 3(a)]. The 18 cm$^{-1}$ oscillator contributes only about 0.5 at $T < T_3$, significantly weakens at $T_2$, and disappears above $T_1$. The difference in the temperature of the incommensurate-commensurate $T_3$ transition obtained from IR and dielectric measurements [compare Figs. 3(a) and 3(d)], is due to the thermal hysteresis. Our IR transmission spectra were measured in the warming regime, while dielectric data were taken upon relatively fast cooling. The quantitative discrepancy between $\Delta\varepsilon(T)$ and $\sum S_j(T)$ at $T_3$ can be influenced by systematic errors of experiment, such as oscillator overdamping due to the sample thickness and imperfections of linear polarizers. Note that the magnitude of $\varepsilon_\infty$ in the parent compound of $TbMn_2O_5$ has been recently related to the contribution of transverse optical phonons,[7] which should be the case for our $HoMn_2O_5$ crystals as well.

Figure 4 shows temperature dependence for the frequency of IR excitations. No strong variations that exceeded the experimental accuracy of the frequency measurement have been detected at the phase transitions. A natural softening of the 97 cm$^{-1}$ excitation is detected at high temperatures possibly due to interaction with low-frequency optical phonons. Fig. 5 represents schematics for the most of the observed optical excitations. Temperature-independent excitation frequencies (for $T \leq 70$ K) are represented with horizontal lines. Dashed, dash-dot, and solid lines correspond to the light polarization along *a*, *b*, and *c* directions, respectively. All observed IR excitations (except the one at 18 cm$^{-1}$) remain optically active in at least one polarization configuration in the paramagnetic phase, *i.e.*, at the temperatures well above $T_1 \approx 43$ K (see Figure 4). Analysis of the results shown in Fig. 5 is given in the next section.

## III. DISCUSSION

$Ho^{3+}$ in $HoMn_2O_5$ has ten 4*f* electrons. According to the Hund's rule, the ground state term for these ten electrons is $^5I_8$.[13] In $HoMn_2O_5$ crystals, 17 states with *J*=8 for Ho 4*f* electrons split due to the ligand field, which, in particular, includes the Hamiltonian term associated with the hybridization of Ho 4*f* and Mn 3*d* orbitals, or the covalent bonding term. We distinguish the ligand field (LF) from the crystal field, to emphasize that the effect of Mn ions on $Ho^{3+}$ is more complex than the Coulomb potential from point charges. Such hybridization between $4f$ states on $Ho^{3+}$ and $3d$ states on Mn, as occurred, for example, in heavy fermions compounds, is expected to give rise to coupling between spin states of Mn ions and ligand field states of Ho ions. The role of intervening oxygen ions cannot be neglected in this consideration and the magnetic interaction between Ho and Mn ions can be considered as the superexchange through the $2p$ orbitals of oxygen.

Several neutron scattering studies of related compounds, such as $HoVO_4$ and $HoMnO_3$, provide us with an estimate for the frequency range of the LF transitions to be between 12 cm$^{-1}$ and 300 cm$^{-1}$.[14,15] For example, the low-energy dispersionless excitations, which have been attributed to the crystal field splitting of $Ho^{3+}$ in $HoMnO_3$, are at 12 and 25 cm$^{-1}$. Therefore, we believe that the IR excitations observed in this work for the frequency range up to 100 cm$^{-1}$ in $HoMn_2O_5$ are mostly, with some possible exceptions, dipolar transitions between the ground state and the low energy excited LF states within the $J = 8$ multiplets.

For electrons in a centro-symmetric potential, electric dipole optical transitions between the levels of the $4f^n$ configuration are all forbidden by the parity rule. In contrast, magnetic dipole transitions are allowed between states of the same parity. The standard angular momentum method (free ion, $LS$ coupling) gives the selection rules for magnetic dipole transitions, which are independent of the crystal field. Magnetic dipoles are active between states with $\Delta J = 0, \pm 1$, where $\Delta J = \pm 1$ corresponds to transitions between the neighboring multiplet components with $\Delta S = 0$, $\Delta L = 0$ and $\Delta J = 0$ is meaningful in external magnetic field only. Although this is always true for a free ion, the lower symmetry of the LF potential in crystals can result in appearance of forced electric dipole transitions in addition to the magnetic dipole transitions. A forced electric dipole transition occurs when $Ho^{3+}$ is placed in a LF that has no center of symmetry, like in $RMn_2O_5$.[16,17] Hence, the wavefunction of $Ho^{3+}$ has a mixed parity. The oscillator strength for the forced electric-dipole transition is determined by the admixture of the states with the opposite parity (*e.g.*, *d* electrons) to the predominantly even parity of $Ho^{3+}$ in the $4f^{10}$ configuration. Selection rules for the forced electric dipole transitions depend on the exact symmetry of the LF and their analysis at the phase transitions in $RMn_2O_5$ requires separate theoretical studies. Here we will only mention that the selection rules for the forced electric dipole transitions with $\Delta J = 0, \pm 1$ require reversed light polarization (circular $\sigma$ vs. linear $\pi$) compared to that for the corresponding magnetic dipoles.[17,18]

According to the Wigner-Eckart theorem,[19] the *z*-direction dipolar transition between two LF states can be non-zero, only if the two states have common $J_z$ components. Similarly, the *x*- or *y*-directional dipolar transitions can be non-zero, only if the two states have $J_z$ components that are different by $\pm 1$. For $HoMn_2O_5$, we choose the *z*-axis to be parallel to the crystalline *c*-axis. Correspondingly, *x*- and *y*-axis are along crystalline *a* and *b* directions. Figure 2 shows that the main change at the magnetic ordering temperatures occurs to the polarization of LF transitions. For instance, the IR excitations at 10.5 cm$^{-1}$ and 13 cm$^{-1}$ are present only for the incident light polarized along *c* direction and are absent for the incident light polarized along *a* and *b* directions for the temperature range above $T_3$. This means that, the ground state and the excited states at 10.5 cm$^{-1}$ and 13 cm$^{-1}$ share the same $J_z$ components, but do not have any $J_z$ components separated by ±1 that do not cancel the contribution of each other to the dipole transition matrix element. The appearance of IR excitations at 10.5 cm$^{-1}$ and 13 cm$^{-1}$ polarized along $a$ and $b$ axes below $T_3$ can be interpreted as the symmetry change of either the ground state or the excited state in such a way that the ground and the excited states have uncompensated $J_z$ components separated by ±1. Such changes can occur, for example, through the mixing with another low energy state, which can be excited from the ground state by $a$ and $b$ polarizations above $T_3$, such as the one at 37 cm$^{-1}$.

Modification in the ligand field, which causes such mixing of electron states below $T_3$, coincides with the commensurate-to-incommensurate AFM phase transition for Mn spins. Since the primary cause of the Ho$^{3+}$ LF splitting is still the electrostatic potential from the surrounding ions, *i. e.*, the crystal field effect, the *energy* spectrum of the LF states is determined by the electrostatic interaction within the crystal lattice. This energy scale stays practically unchanged in a wide temperature range. However, magnetic effects at $T < T_1$, which appear due to the mixing (or covalent bonding) of Ho$^{3+}$ 4$f$ states with the ordered spins of the nearest neighbor Mn 3$d$ states, influences the *symmetry* of LF states, which can change at each magnetic ordering temperature resulting in transformation of the selection rules for the corresponding optical transitions. The weak nature of 4$f$–3$d$ mixing does not allow significant shifts of the level *energy* (see Fig. 4), but affects in the first order of the perturbation theory the *selection rules* for optical transitions. Thus, our results imply that one of the main causes of the sudden steps in the dielectric constant along *a* and *b* directions at $T_3$ can be attributed to the contribution of the forced electric dipole transitions between the $Ho^{3+}$ LF states due to their coupling to the Mn sub-lattice. Similar energy scale between magnetic interaction and LF splitting supports the importance of the LF splitting connecting magnetism and dielectric properties of the multiferroic $HoMn_2O_5$ and, probably, of other related $RMn_2O_5$ compounds with $R$ = Tb and Dy.

In addition to the coupling between $R^{3+}$ LF states and the magnetism on the Mn sub-lattice, of the special interest are the symmetry changes due to magnetically induced lattice modulation and dynamic interactions with the lattice. The latter is especially important for the case when the high-energy LF states are in resonance with the lower-frequency IR-active TO phonons (~150 cm$^{-1}$), but this mechanism should not have any strong temperature dependence. In contrast, the influence of magneto-striction on the change of selection rules for the LF transitions can result in anomalies at the phase transitions. However, the strongest changes in the lattice distortion for $RMn_2O_5$ compounds are usually observed between $T_3$ and $T_1$.[20] Since in our experiments the change of selection rules for the LF transitions occurs for $T<T_3$ we consider this as a supporting argument for our interpretation based on the coupling between the magnetism on the Mn sub-lattice and $Ho^{3+}$ LF states. The splitting of the 12 cm$^{-1}$ state into a doublet (10.5 and 13 cm-1) at $T_C = T_2 = 39$ K can be attributed to the symmetry lowering, which is caused by electric polarization.

The coupling between Mn spin and Ho$^{3+}$ LF state opens the possibility of composite excitations of magnon and LF states in this compound, particularly due to their similar energy scale. Two main criteria of identification of such composite excitations should be (i) their activity only in the temperature range below $T_1$ and (ii) a nonzero dispersion, which is typical for AFM magnons. Can we identify such excitations in our spectra for HoMn$_2$O$_5$? The only candidate for the composite LF-magnon is the peak at 18 cm$^{-1}$, which is close to the LF transitions at 10.5 cm$^{-1}$ and 13 cm$^{-1}$. This peak dominates in the *(b,b)* polarization below $T_3$, maintains sufficient strength above $T_3$, but absent for all polarizations above $T_1$, thus satisfying the first criterion for electric-dipole active LF-magnon excitations. But since the second criterion cannot be verified with the IR transmission technique, our interpretation requires additional experimental proofs, which can be obtained, for example, using neutron scattering. Thus, the question whether the IR transition at 18 cm$^{-1}$ is a composite LF-magnon excitation or just another LF state which became electric-dipole-active due to the symmetry change below $T_1$ remains open. In any case, however, we emphasize that the oscillator strength of the excitation at 18 cm$^{-1}$ is only 20% of the combined oscillator strength for the low-frequency LF doublet at 10.5 and 13 cm$^{-1}$, thus making this proposed LF-magnon of a secondary importance for explanation of the step-like behavior of the static dielectric function. Appearance of this excitation in *(a,a)* polarization with a significantly lower oscillator strength can be interpreted as due to a mixed electric- and magnetic-dipole activity of this excitation.

We can also identify excitations from thermally excited LF states (31, 80, and 85 cm$^{-1}$). The temperature-activated transitions at 31 and 85 cm$^{-1}$ are marked with arrows in Fig. 6(a), which compares two transmission spectra measured at $T = 5$ and 18 K. The absorption peak at 85 cm$^{-1}$ appears below $T_3$ and becomes stronger as the temperature increases. This IR excitation can be considered as the optical transition from the lowest excited LF state (the doublet at 10.5 and 13 cm$^{-1}$) to the higher energy LF state at 97 cm$^{-1}$. Similarly, the 31 cm$^{-1}$ excitation below $T_3$ can be considered as the excitation from the same thermally excited LF state to the 42 cm$^{-1}$ excited state. These thermally excited excitations can be found in the temperature-frequency maps in Fig. 2(a) and they are also indicated in Fig. 5 with dashed vertical arrows. For the thermally activated transitions between two excited LF states, $|1\rangle$ and $|2\rangle$, one can expect that the transition probability $W_{|1\rangle \to |2\rangle}$ depends on temperature-activated occupation of the $|1\rangle$ excited state. At low temperatures, this dependence is dominated by the quasi-linear in $T$ term, while at higher temperatures (above 60 K for 85 cm$^{-1}$ transition), depopulation of the ground state to other excited states results in a slow decrease of transition probability. Figure 6(b) compares the probability of the thermally-activated transitions $W_{|1\rangle \to |2\rangle}(T)$, which has been calculated in a simplified six-level model, with experimental data for the oscillator strength of 85 cm$^{-1}$ excitation. Exact calculations should take into account the position and wavefunction symmetry for all 4$f$ Ho$^{3+}$ states, which should give a better agreement with experiment.

Our IR transmission spectra did not reveal any sharp variations at $T_C = T_2$ and we cannot assign any specific oscillator that changes its strength or frequency explaining the spike in $\varepsilon(T)$ at $T_2$ shown in Figs. 3(d,e). The corresponding feature of the static dielectric function in TbMn$_2$O$_5$ was explained in Ref. [10] by the zero-frequency electromagnon mode. Here we suggest that this peak is related to the spontaneous electric polarization due to the ferroelectric domain structure.

Far-IR transmission spectra in $R$Mn$_2$O$_5$ compounds vary for different elements, like Tb, Dy, and Ho. Both, the lowest IR transition frequency and the higher frequency absorption peaks, change for different compounds by a few cm$^{-1}$. Table 1 includes frequencies of LF transitions in HoMn$_2$O$_5$ along with the lowest optical transition frequency in TbMn$_2$O$_5$, which was attributed to an "electromagnon" in Ref [10]. In contrast to the transmission results for TbMn$_2$O$_5$ with a strong "electromagnon" mode at 9.6 cm$^{-1}$ polarized along $b$-axis, in HoMn$_2$O$_5$ we have observed a LF doublet at 10.5 and 13 cm$^{-1}$. As it was described above, this transition is IR-active in both $a(c,c)$ and $a(b,b)$ configurations being preferably polarized *perpendicular* to the direction of spontaneous polarization $b$ with the polarization degree of $(S^{a(c,c)} - S^{a(b,b)})/(S^{a(c,c)} + S^{a(b,b)}) = 0.4$. Here $S^{a(c,c)}$ and $S^{a(b,b)}$ are the oscillator strengths from Table 1 for $a(c,c)$ and $a(b,b)$ configurations, respectively.

We further comment on "electromagnon", a new composite excitation recently proposed for $R$MnO$_3$ ($R$ = Gd and Tb)[8] and $R$Mn$_2$O$_5$ ($R$ = Y and Tb).[10] The origin of electric-dipole activity for this excitation with a typical frequency between 10 and 20 cm$^{-1}$ was attributed in Refs. [9,10] to the interaction with optical phonons. In spite of the strong spin-lattice coupling in these multiferroic compounds, it is still under debate whether the two excitations separated by a large energy scale (optical phonon having usually one order of magnitude larger energy than "electromagnons"), could experience a direct coupling, which is strong enough to explain magneto-electric effect. In compounds like $R$Mn$_2$O$_5$ and $R$MnO$_3$ with $R$ = Gd, Tb, Dy, and Ho (rare-earth ions with *f*-electrons) it might be more fruitful to consider composite excitations between magnons and forced electric-dipole LF excitations since they have a more comparable energy scales. In general case, dynamic interactions between the lattice and magnons can be mediated by the LF states, which have resonances in both, magnon and phonon spectral ranges. For other multiferroic compounds with rare-earth ions without nominal *f*-electrons, the physics can be even more versatile and the concept of "electromagnons" could be more valuable, especially in the case of YMn$_2$O$_5$ compound[10] with a direct *3d-3d* interaction.

## IV. CONCLUSIONS

Our experiments demonstrated a strong influence of the ligand field between Ho$^{3+}$ and Mn spin states on IR transmission properties of HoMn$_2$O$_5$. This coupling can explain the step-like anomalies in the dielectric constant in $R$Mn$_2$O$_5$ compounds with rare-earth ions having incomplete 4*f*-shell. Further magneto-optical and neutron spectroscopy experiments are required to confirm the existence of LF-magnon composite excitations.

## ACKNOWLEDGEMENTS

Authors are thankful to C. Bernhard, Dennis Drew, V. Kiryukhin, L. Mihály, and T. Zhou for valuables discussions and to R. Smith for help at U12IR beamline. Work at NJIT was supported by the NSF-DMR-0546985. Work at Rutgers was supported by the NSF-DMR-0405682. Use of the National Synchrotron Light Source, Brookhaven

National Laboratory, was supported by the U.S. Department of Energy, Office of Science, Office of Basic Energy Sciences, under Contract No. DE-AC02-98CH10886.18

TABLE I. Polarization, frequency $\Omega_j$, and oscillator strength $S_j$ of the low-frequency ligand field optical transitions in $HoMn_2O_5$. The units of the oscillator strength $S_j$ are the same as that for the static dielectric function. The type of electric dipole, magnetic dipole, or mixed activity is marked with ED, MD, and ED+MD, respectively.

| | Polarization | $\Omega$, cm$^{-1}$ | $<S_j>$ $T = 5-19$ K | Type | Comments |
|---|---|---|---|---|---|
| $HoMn_2O_5$ | $a(b,b)$ | 10.5 and 13 | 1.3 | ED | Disappears above $T_3$ |
| | | 18 | 0.6 | ED+MD | Disappears above $T_2$ |
| | | 37 | | ED+MD | Disappears above 200 K |
| | | 42 | | ED+MD | Weak; Disappears above $T_3$ |
| | | 97 | | MD | |
| | $b(a,a)$ | 10.5 and 13 | 0.5 | ED | Disappears above $T_3$ |
| | | 18 | 0.3 | ED+MD | Disappears above $T_3$ |
| | | 37 | | ED+MD | Disappears above 150 K |
| | | 97 | | MD | |
| | $a(c,c)$ and $b(c,c)$ | 10.5 and 13 | 3 | ED | Disappears above 150 K |
| | | 37 | | ED+MD | Disappears above $T_2$ |
| | | 42 | | ED+MD | Disappears above 150 K |
| | | 52 | | ED | Weak |
| $TbMn_2O_5$ Ref. 10 | $\vec{e} \parallel b$ | 9.6 | 3.6 | ED | "electromagnon" Ref. 10 |

**FIGIRE CAPTIONS:**

FIG. 1 (color online). (a), (b), (c) Normalized Fourier-transform far-IR transmission spectra of HoMn$_2$O$_5$ single crystal measured at $T = 5$ K in $a(b,b)$, $a(c,c)$, and $b(a,a)$ configurations, respectively. Arrows indicate the frequencies of IR-active absorption lines.

FIG. 2 (color online). Maps of transmitted intensity *vs.* temperature and frequency for HoMn$_2$O$_5$ in (a) $a(b,b)$, (b) $a(c,c)$, and (c) $b(a,a)$ configurations. Red (dark) color corresponds to stronger absorption, while blue (light) color indicates high transmission. The scales of the transmission intensity for each plot are the same as that for the corresponding graphs in Figure 1.

FIG. 3 (color online). (a), (b), and (c) Temperature dependence of the oscillator strength $S_j$ presented in units of the static dielectric function $\varepsilon(0)$ in $a(b,b)$, $b(a,a)$, and $a(c,c)$ polarization configurations, respectively. (d), (e), and (f) Temperature dependencies of $\varepsilon_a$, $\varepsilon_b$, and $\varepsilon_c$ measured for different orientations of the same HoMn$_2$O$_5$ crystals. The shift of the $T_3$ transition temperature between the IR and dielectric measurements is shown with horizontal arrows in (d) and (e).

**FIG. 4** (color online). Temperature dependence for the IR oscillator frequencies measured in two polarizations: *a(b,b)* and *a(c,c)*. The error bars for the 80 cm$^{-1}$ absorption line indicate an increase of the experimental uncertainty at high temperatures.

**FIG. 5** (color online). Schematics of the energy diagram for optical transitions between ligand field split levels of the Ho$^{3+}$ multiplet ($^5I_8$) in HoMn$_2$O$_5$. The energies of experimentally observed transitions from the ground state are represented with horizontal lines. Dashed, dash-dot, and solid lines indicate the allowed polarization for the corresponding transitions: *a(b,b)*, *b(a,a)*, and *a(c,c)*, respectively. Short vertical arrows indicate thermal activation from the ground state to the nearest excited state with a consecutive optical transition (31 and 85 cm$^{-1}$) to the higher energy states. Splitting of the lowest excited state (12 cm$^{-1}$ → 10.5 + 13 cm$^{-1}$) occurs in the vicinity of $T_2$ upon cooling. Vertical dash-dot lines divide different magnetic phases: incommensurate AFM (IC), commensurate AFM (C), and paramagnetic (PM).

**FIG. 6** (color online). (a) Normalized Fourier-transform far-IR transmission spectra of a HoMn$_2$O$_5$ single crystal measured at $T$ = 5 K and 18 K in *a(b,b)* configurations. Arrows indicate the frequencies of the temperature-activated optical transitions at 31 and 85 cm$^{-1}$. (b) Temperature dependence of the oscillator strength for 85 cm$^{-1}$ transition (circles). Solid curve (right scale) shows the expected trend for the probability of the temperature-activated transition between two excited LF states, $|1\rangle$ at 12 and $|2\rangle$ at 97 cm$^{-1}$, which was calculated in a simplified model for a 6-level system.

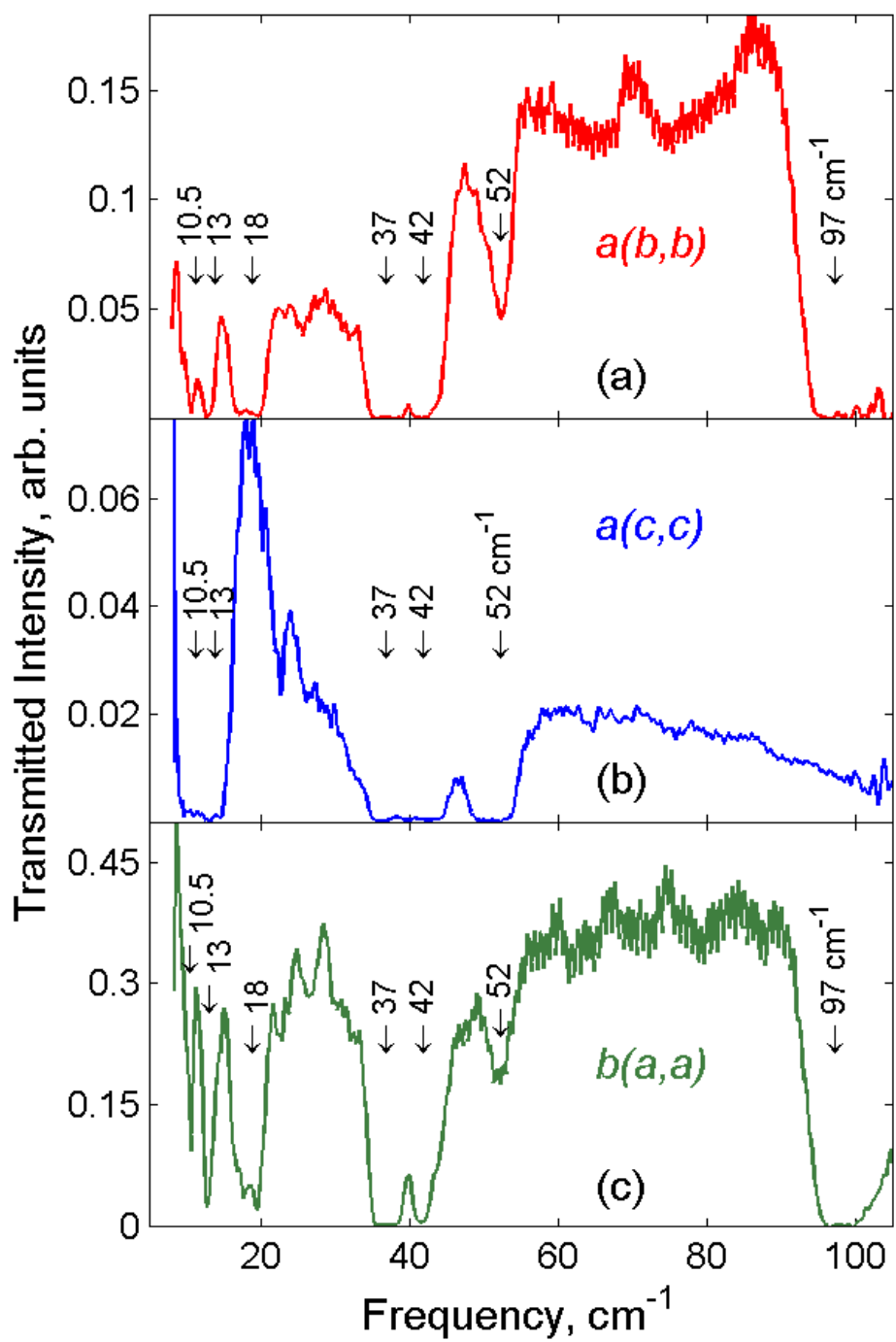

FIG. 1

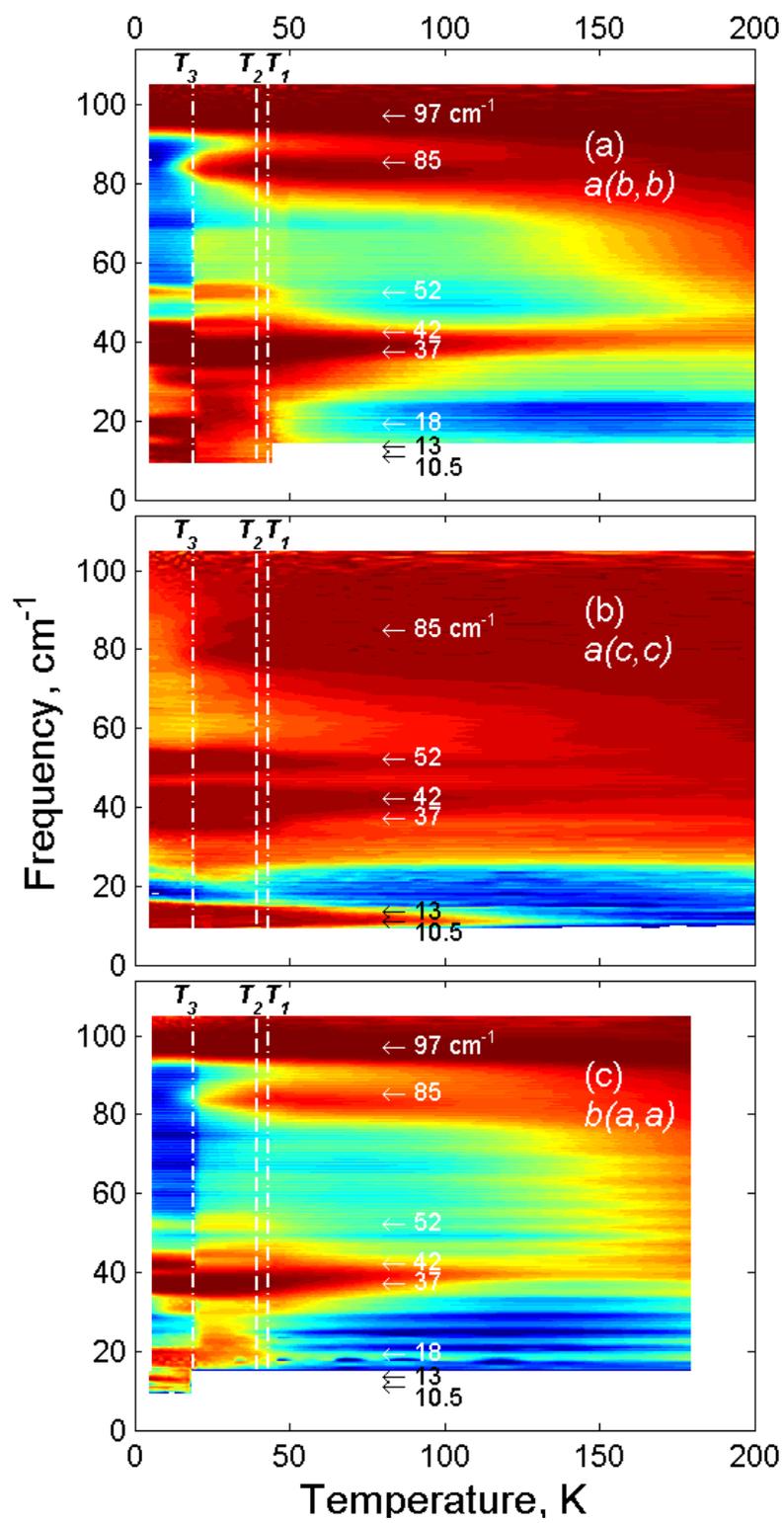

FIG. 2

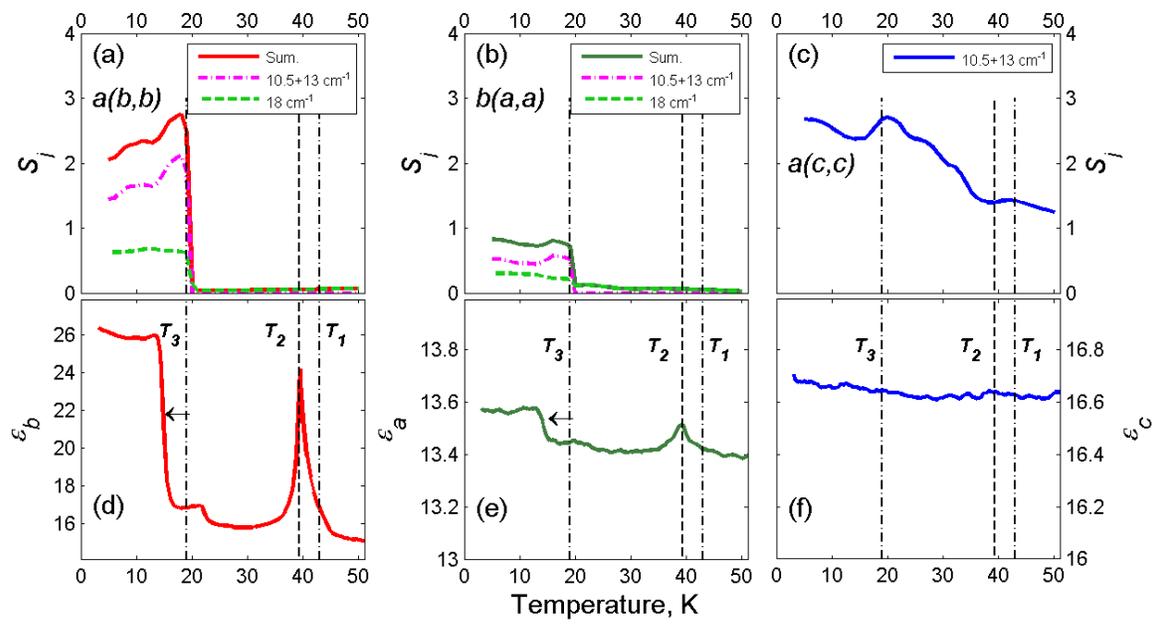

FIG. 3

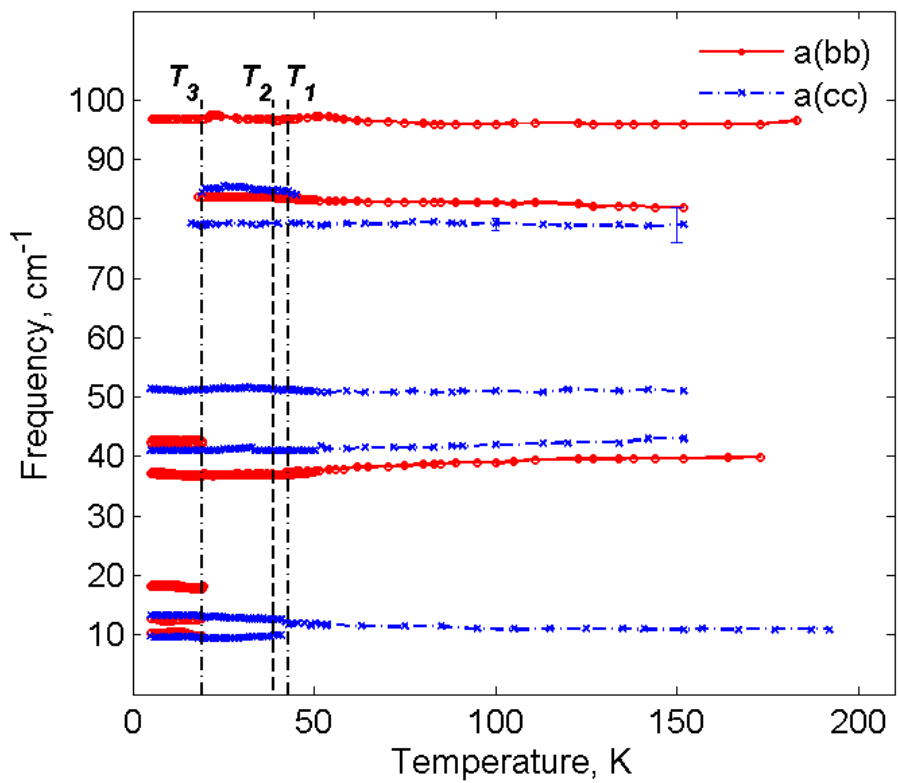

FIG. 4

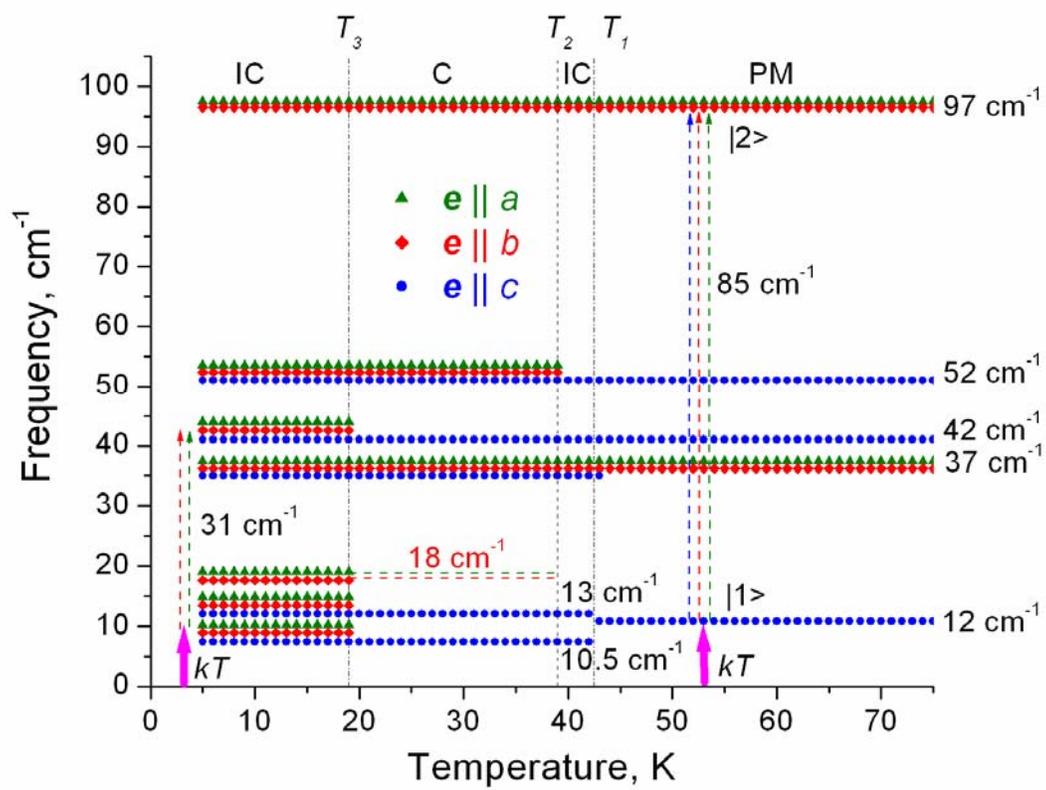

FIG. 5

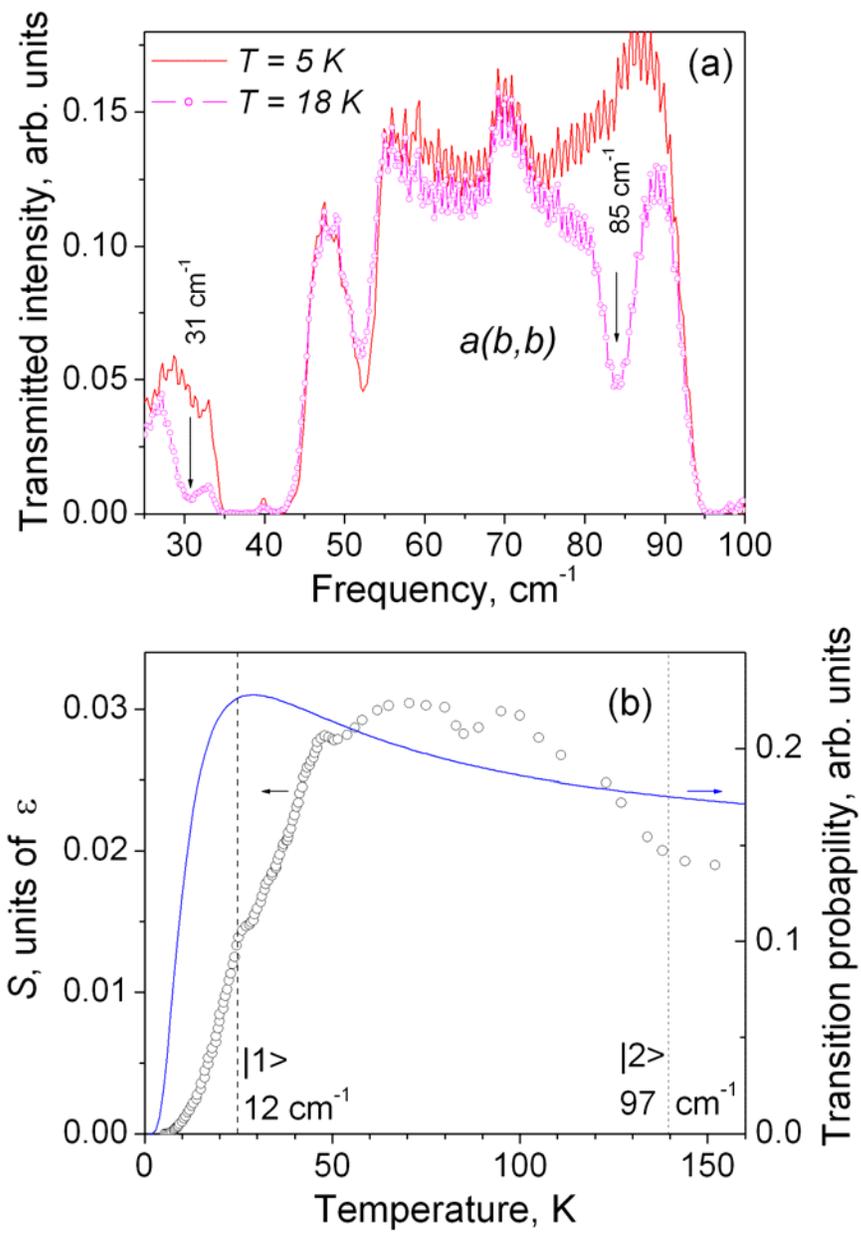

FIG. 6